\documentclass[twocolumn,showpacs]{revtex4}
\usepackage{graphicx,subfigure,color}
\usepackage{bm,xspace}
\makeatletter

\begin{document}

% Macros for Dirac bra-ket <|> notation
\def\bra#1{\mathinner{\langle{#1}|}}
\def\ket#1{\mathinner{|{#1}\rangle}}
\def\braket#1{\mathinner{\langle{#1}\rangle}}
\def\ave#1{\mathinner{\langle{#1}\rangle}}

\preprint{This line only printed with preprint option}
%\title{Controllable spontaneous emission  of quantum dot in planar photonic crystal waveguide-cavity system}
\title{Nonlinear photoluminescence spectra from a quantum dot-cavity system: Direct evidence of
pump-induced stimulated emission and anharmonic cavity-QED}
\author{Peijun Yao}
\author{P. K. Pathak}
\author{E. Illes}
\author{S. Hughes}
%\email{shughes@physics.queensu.ca}
\address{Department of Physics, Queen's University\\
Kingston, ON  K7L 3N6 Canada}
\author{S. M\"unch, S. Reitzenstein, P. Franeck, A. L\"offler,
T. Heindel, S. H\"ofling, L. Worschech, and A. Forchel}
\address{Technische Physik, Physikalisches Institut,
Universit\"at W\"urzburg and Wilhelm Conrad R\"ontgen Research Center for Complex Material Systems,
Am Hubland, D-97074 W\"urzburg, Germany}
%\email{shughes@physics.queensu.ca}
\begin{abstract}

We
%theoretically
investigate the power-dependent photoluminescence spectra from a strongly coupled quantum dot-cavity system
using a quantum master equation  technique that accounts for
incoherent pumping,
pure dephasing, and fermion or boson statistics.
Analytical spectra at the one-photon correlation level and the numerically exact
multi-photon spectra for fermions are presented.
We compare
to recent experiments on a quantum dot-micropiller cavity system and show
%, in contrast to previous
%theoretical formalisms,
that an excellent fit to the data can be obtained by varying only the incoherent pump rates
in direct correspondence with the experiments.
%Moreover,
%%In particular,
%we explore the coupled QD-microcavity system for very high pump rates,
%and demonstrate the profound influence of stimulated emission.
%%, otherwise
% %was not properly described by previous theoretical formalisms which predict for instance
%%  negative mean photon numbers will be predicted.
Our theory and experiments together
show a clear and systematic way of studying stimulated-emission induced
broadening and anharmonic cavity-QED.
%We also show that while the mean photon number can easily be greater than 1,
%single exciton lasing.
%, a regime that is in part forbidden by an incoherent cavity pump.
%, and the striking influence
%that stimulated emission has on the power-dependent spectra.
%and  analyzing  single exciton lasing  photon correlations.
\end{abstract}

%\pacs{****}
%42.50.Ct Light interaction with matter
% %78.67.Hc:quantum dot,
%32.70.Jz:Line shapes, widths, and shifts
%42.50.Pq: Cavity quantum electrodynamics,

\maketitle

%\section{Introduction}

{\em Introduction.--} Single  quantum dot (QD) - cavity systems
facilitate  the realization of  solid state qubits (quantum bits) and have applications
for producing single photons~\cite{michler,moreau,santori2}
and entangled photons~\cite{ellis_njp043035,johne}.
Rich in physics and potential applications,
the coupled QD-cavity has been  inspiring  theoretical and experimental
groups to probe deeper into the underlying physics of both weak
and strong coupling regimes of semiconductor cavity-QED (quantum electrodynamics).
Key signatures of cavity-QED include the Purcell effect and  {\em vacuum} Rabi oscillations.
Although a well known phenomenon in atomic cavity optics~\cite{McKeeverPRL2004},
vacuum Rabi splitting in a semiconductor
 cavity was only realized a few years ago~\cite{reithmaier_nature432,yoshie_nature432,peterPRL2005}.
%A  common feature of these
%  semiconductor
%  cavities
% structures
% is that they have suitably large $Q/V_{\rm eff}$ ratios --
% with $Q$ the quality factor
%and $V_{\rm eff}$ the effective mode volume -- to allow single  electron-hole pair (exciton) strong coupling.
Inspired by the recent surge of related experiments,
many researchers have been working hard to develop new
theoretical tools to understand the semiconductor cavity-QED systems.
For example,
%in the regime of off-resonant cavity coupling,
the persistent excitation of the cavity mode for large exciton-cavity detunings
was measured~\cite{hennessy_nature445,press_PRL07}, and qualitatively
explained by extended theoretical approaches that
account for coupling between the {\em leaky} cavity mode and the exciton, and by
showing that the main contribution to the emitted spectrum comes from the
cavity-mode emission~\cite{Raymer2006,Auffeves2008,Yamaguchi2008,Naesby2008,hughesOE2009}.
These formalisms assume an initially excited exciton or an initially excited leaky cavity mode, and they are valid
for low pump powers. However, an interesting question that has been posed recently, e.g., see Refs.~\cite{Keldysh2006,laussyPRL2008,laussyPRB2009}, is what is the role of an incoherent pump on the photoluminescence (PL) spectra, where the pump can excite
the exciton or cavity mode?
To experimentally investigate the pump-dependent spectra,
two recent experiments have been
respectively reported  by M\"unch {\em et al.}~\cite{munch_oe12821}
for a QD-micropillar system, and
 by Laucht {\em et al.}~\cite{arnePRL2009}
 for a QD-photonic crystal system;
 these measurements show the  pump-induced crossover from strong to weak coupling.
 % as a function of pump power.
%Although qualitatively similar experiments, one data set was decribed in terms of a boson
%model and the other was described in terms of a fermion model. Both models

In this work, we present a master equation (ME) theory
that self-consistently includes incoherent pumping, stimulated emission, and pure dephasing.
% and
%stimulated emission of the incoherent cavity pump.
%
%, pure dephasing, and
%concohernet pumping from
%extend previous semiconductor cavity-QED theories~\cite{laussyPRL2008,laussyPRB2009,arnePRL2009}
%by introducing a model that self-consistently includes
%stimulated emission.
We derive analytical results
at the level of one-photon correlations and present {\em numerically exact}  results
for the multi-photon spectra.
 We analyze the
W\"urzburg~\cite{munch_oe12821}
experiments directly and show the striking differences with previous
models that neglect the direct influence of stimulated emission~\cite{laussyPRL2008,laussyPRB2009,arnePRL2009}.
For the incoherent pumping of the exciton, we present two
ME models:
a thermal bath model (c.f.~a two-level system) and a heat bath
model at large negative temperature (c.f.~a multi-level laser system).
% the latter model has also been reported very
%recently by Ridolfo {\em et al.}~\cite{ridolfo_arxiv}.
{\em Accounting for fermion statistics, pure dephasing, and the thermal bath model,
% for exciton pumping,
an excellent fit to the data is obtained by  only changing the
incoherent pump rates in direct correspondence with the experiments}.
%We also discuss the possible regime of single exciton lasing.

%\bibitem{ridolfo_arxiv}  A. Ridolfo {\em et al.},
%% O. Di Stefano,  S. Portolan,  and S. Savasta,
% %Photoluminescence from Microcavities Strongly Coupled to Single Quantum Dots, 	
% arXiv:0906.1455v1 [cond-mat.mes-hall].

\begin{figure}[!t]
%\hspace{-0.4cm}
%\includegraphics[width=0.46\textwidth, height=0.36\textwidth]{fig1.eps}
\includegraphics[width=0.46\textwidth]{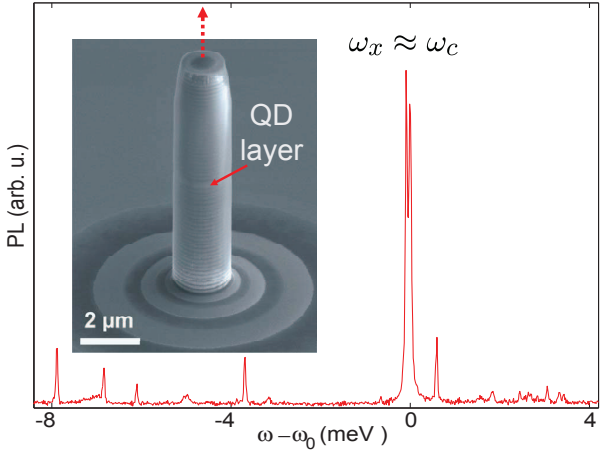}
% Here is how to import EPS art
\vspace{-0.3cm}
\caption{(Color online) Typical broadband PL spectrum that is emitted when a
target exciton  is closely resonant  with the cavity mode (near $\omega_0 = 1331.355\,$meV); away from the target
exciton, there are
%clearly
a series of other exciton levels that can also
couple, off-resonantly, to the cavity mode.
The SEM image shows our micropillar cavity and the  QD layer. The emitted
photons from the QDs are detected through vertical emission.}
\label{fig:schematic}
\end{figure}

{\em Cavity system and model.--} The system investigated here is shown
as a scanning electron microscope (SEM) image  in Fig.~\ref{fig:schematic}, along
with the extended experimental data
of Ref.~\cite{munch_oe12821}.
We make the following assumptions:
 the cavity is single-mode in the frequency of interest; the coupling between the cavity and target QD exciton is
 described through a coupling rate $g$;  the decay rate of cavity is $\Gamma_c$;
 for the strongly coupled QD, we include only the target exciton as a system operator, and consider
 both  radiative decay, $\Gamma_x$, and pure dephasing, $\Gamma_{x}'$.
 The QD-cavity system is driven simultaneously by an exciton pump, $P_x$, and a cavity pump, $P_c$; the former is caused by the {\em incoherent} relaxation of electron-hole pairs from the higher energy level, and the latter is due to the cavity coupling with off-resonant excitons
  (probably coming from other QDs  in the cavity layer).
%signatures of   these {\em background emitters}, that may feed the cavity mode, are
%  visible in  Fig.~1.
 To treat the {\em incoherent} excitation, we consider
   a system-reservoir interaction~\cite{carmichael_book},
   apply a Born-Markov approximation, and trace over the cavity and
   target exciton pump reservoirs ({\em bath approximation}). We have
\begin{eqnarray}
\frac{d\rho}{dt}=\frac{-i}{\hbar}[H_s, \rho]+  {\cal L}(\rho),
\end{eqnarray}
with the system Hamiltonian, $H_s=\hbar \omega_x {\hat{\sigma}}^+{\hat{\sigma}}^-+\hbar \omega_c{\hat{a}}^\dagger{\hat{a}} \ +\hbar g({\hat{\sigma}}^-\hat{a}^\dagger + \hat{\sigma}^+\hat{a})$, where
$\hat a $ represents the cavity mode operator,
\textbf{$\hat \sigma^{+/-}$} are the Pauli operators of the target QD
exciton (with resonance frequency   $\omega_x$), and $\omega_c$ is the
eigenfrequency of the leaky cavity mode.
The target exciton  and cavity mode  get pumped incoherently through the corresponding reservoirs.
The state of the reservoirs
can be written as $\rho_P^O=\sum_k \rho_{kk}^O\ket{n_k^O}\bra{n_k^O}$, for $O=x,c$; where $\rho_{kk}^O$ is the density of
reservoir modes and $n_k^O$ is number of photons in the mode of wave vector k.
The correlations for the photon reservoir operators $\hat a_k^c$ are given by $\braket{\hat a_k^c}=0$,
$\braket{(\hat a^{c}_k)^\dagger \hat a_{k'}^c}=\bar n_k^c\delta_{kk'}$ , and
$\braket{\hat a^c_k (\hat a_k)^\dagger}=(\bar n_k^c+1)\delta_{kk'}$. Defining the average pump photon number around the cavity frequency as $\bar n^c=\bar n_k^c$, at $k=\omega_c/c$,
yields the effective  incoherent cavity pump rate:  $P_c=\Gamma_c \, \bar n^c$.
This incoherent pump process
%is well known
%, e.g., see
% is an appropriate model for a
%bosonic pump, and, e.g.,
agrees with the model
of
 Tian and Carmichael~\cite{carmichaelQO1992}.
The superoperator in Eq.~(1) becomes
\begin{eqnarray}
&&{\cal L (\rho)}  =
\frac{P_c}{2}
\left ( 2 \hat{a}^\dagger \rho \hat{a} - \hat{a} \hat{a}^\dagger \rho - \rho \hat{a} \hat{a}^\dagger
\right ),   \nonumber \\
&\hbox{}  + & \!\!\! \frac{\Gamma_c+P_c}{2}
\left ( 2 a \rho \hat{a}^\dagger - \hat{a}^\dagger \hat{a} \rho- \rho \hat{a}^\dagger \hat{a}
\right )  , \nonumber \\
&\hbox{} + &  \!\!\! \frac{P_{12}}{2}
\left ( 2 \hat{\sigma}^+ \rho \hat{\sigma}^-  - \hat{\sigma}^-\hat{\sigma}^+\rho - \rho \hat{\sigma}^-\hat{\sigma}^+
\right )  \nonumber \\
&\hbox{} + &  \!\!\!\frac{P_{21}}{2}
\left ( 2 \hat{\sigma}^- \rho\hat{\sigma}^+ -\hat{\sigma}^+\hat{\sigma}^-\rho - \rho \hat{\sigma}^+\hat{\sigma}^-
\right ) \! +\!\frac{\Gamma_{x}'}{4}
\left ( \hat{\sigma}_z\rho \hat{\sigma}_z \!- \! \rho
\right ),
   \ \ \ \ \
\end{eqnarray}
which is
in
Lindblad form.
For the  exciton pump we consider two different models:
{\em model-1} (thermal bath):  $P_{12} = P_x$ and $P_{21}=\Gamma_x+P_x$;
{\em model-2} (heat bath at large negative temperatures)~\cite{LofflerPRA1997}: $P_{12} = P_x$ and $P_{21}=\Gamma_x$;
$P_x$ is the target exciton pump rate which
%, which,
%like $P_c$,
is presumed to proportionally follow the experimental pump power.

%Naturally, only the inversion model allows
%lasing.
%, and is likely not a good model for modeling
%low pump power quantum statistics.
%The previous MEs  used explain pump-dependent PL in semiconductor %cavities~\cite{laussyPRL2008,arnePRL2009},
%have the $\Gamma_c + P_c$ term above replaced by $\Gamma_c$.
%; thus,
%stimulated emission  was neglected.
% Although one could argue that such terms are
% partly accounted for by assuming an effective $P_c- \Gamma_c$, such assumptions
%challenge our understanding of the underlying physics.
%especially for high pump rates, where
%$P_c > \Gamma_c$,
%a regime
%that should
%be accessible experimentally.

%Armed with the above ME,  which includes stimulated emission processes,
One can next derive analytical
spectra at the level of one-photon correlations, or compute the exact numerical spectra for
$n-$photon correlations, e.g., see Refs.~\cite{LofflerPRA1997,laussyPRL2008}).
We will present both approaches.
Using Eq.~(2), adopting the one photon-correlation
  approximation $\ave{\hat{\sigma}_z \hat{a}}=-\ave{a}$,
  %~\cite{one_photon},
and applying
 fermion statistics $[\hat{\sigma}^-, \hat{\sigma}^+]_+=1$,
 we  exploit the quantum regression theorem~\cite{carmichael_book}
to derive the  equation of motion for the
two-time correlation functions,
${d\ave{\hat{a}^\dagger(t) \hat{a}(t+\tau)}}/{d\tau}$
and ${d\ave{\hat{a}^\dagger(t) \hat{\sigma}^-(t+\tau)}}/{d\tau}$.
%%
%%Starting from the above master equation, using the fact $\frac{d\langle{O}\rangle}{dt}=tr\{\rho\dot{O}\}=tr\{\dot\rho{O}\}$ and the rotating property of trace $tr\{ABC\}=tr\{CAB\}=tr\{BCA\}$, we can get a set of equations of $\ave{\hat{\sigma}^-}$, $\ave{\hat{a}}$, here, we adopt the fermionic model which means $[\hat{\sigma}^-, \hat{\sigma}^+]_+=1$ and weak approximation $\ave{\hat{\sigma}_z \hat{a}}=-\ave{a}$. Then we can apply quantum regression theory to get the equation of  corresponding two time correlation function. They are
%${d\ave{\hat{a}^\dagger(t) \hat{a}(t+\tau)}}/{d\tau}=-i(\omega_c-\frac{i}{2}\Gamma_c)\ave{\hat{a}^\dagger(t) \hat{a}(t+\tau)} -ig\ave{\hat{a}^\dagger(t) \hat{\sigma}^-(t+\tau)}\label{eq:}$,
%and
%${d\ave{\hat{a}^\dagger(t) \hat{\sigma}^-(t+\tau)}}/{d\tau}=-i[\omega_x-\frac{i}{2}(2P_x+\Gamma_x+\Gamma_{x}')]
%\ave{\hat{a}^\dagger(t) \hat{\sigma}^-(t+\tau)}-ig\ave{\hat{a}^\dagger(t) \hat{a}(t+\tau)}\label{eq:}$.
%%
%%\begin{eqnarray}
%%\!\!\!\!\!\!&&\frac{d\ave{\hat{a}^\dagger(t) \hat{a}(t+\tau)}}{d\tau}=-i[\omega_c-\frac{i}{2}\Gamma_c]\ave{\hat{a}^\dagger(t) \hat{a}(t+\tau)}\nonumber\\
%%\!\!\!\!\!\!&&\quad\qquad\qquad-ig\ave{\hat{a}^\dagger(t) \hat{\sigma}^-(t+\tau)}\label{eq:} ,\\
%%\!\!\!\!\!\!&&\frac{d\ave{\hat{a}^\dagger(t) \hat{\sigma}^-(t+\tau)}}{d\tau}=-i[\omega_x+\frac{i}{2}(-2P_x-\Gamma_x-\Gamma_{dx})]\nonumber\\
%%\!\!\!\!\!\!&&\quad\qquad\qquad\times\ave{\hat{a}^\dagger(t) \hat{\sigma}^-(t+\tau)}-ig\ave{\hat{a}^\dagger(t) \hat{a}(t+\tau)}\label{eq:}.
%%\end{eqnarray}
Subsequently,
%by
%carrying out a
%Laplace transform,
the steady-state form of the
dominant cavity-emitted spectrum~\cite{hughesOE2009} is obtained
from
$S_{\rm cav}({ R},\omega) = F_{\rm cav}({ R})\, S_{\rm cav}(\omega)$, with
$S_{\rm cav}(\omega)=\Gamma_c/\pi\, {\rm lim}_{t\rightarrow\infty} {\rm Re}\{\int_0^\infty \ave{\hat a^\dagger(t)
\hat a(t+\tau)}
e^{i\omega \tau}d\tau\}$,
where $F_{\rm cav}({ R})$ is a geometrical factor that depends on
the  detector/collection optics.
One obtains
\begin{eqnarray}
S_{\rm cav}(\omega)=\frac{\Gamma_c}{\pi}\, {\rm Re} \left [ \frac{i\ave{\hat{a}^\dagger \hat{a}}_{ss}D(\omega)}{C(\omega)D(\omega)-g^2}+
\frac{ig\ave{\hat{a}^\dagger\hat{\sigma}^-}_{ss}}{C(\omega)D(\omega)-g^2} \right ] , \
\end{eqnarray}
 where $C(\omega)=\omega-\omega_c+\frac{i}{2}\Gamma_c$, and $D(\omega)=\omega-\omega_x+\frac{i}{2}(P_{21}+P_{12}+\Gamma_{x}')$.
The subscript `$ss$' represents the steady-state solutions,
that are given by
\begin{eqnarray}
&&\!\!\!\!\!\!\!\!\!\ave{\hat{a}^\dagger \hat{a}}_{ss}\!\!\!=\!\!\frac{g^2\Gamma_{}(P_{12}\!+\!P_c)\!+\!P_c(P_{21}\!+\!P_{12})
\left(\frac{\Gamma_{}^2}{4}\!+\!\Delta_{cx}^2\right)}
{g^2\Gamma_{}(P_{21}\!+\!P_{12}+\!\Gamma_c) +\Gamma_c
(P_{21}\!+\!P_{12})(\frac{\Gamma_{}^2}{4}\!+\!\Delta_{cx}^2)}, \ \ \ \
\\
&&\!\!\!\!\!\!\!\!\!\ave{\hat{a}^\dagger\hat{\sigma}^-}_{ss}\!\!\!= \!\!\frac{-ig(\ave{\hat{a}^\dagger \hat{a}}_{ss}-\frac{P_{12}}{P_{21}+P_{12}})\left(i\Delta_{cx}+\frac{\Gamma_{}}{2}\right)}{\frac{\Gamma_{}^2}{4}+\Delta_{cx}^2
+\frac{g^2}{P_{21}+P_{12}}\Gamma_{}}, \\
&&\!\!\!\!\!\!\!\!\!\ave{\hat{\sigma}^+\hat{\sigma}^-}_{ss}\!\!\!=\!\!
\frac{P_{12}+ig(\ave{\hat{a}^\dagger\hat{\sigma}^-}_{ss}-\ave{\hat{a}\hat{\sigma}^+}_{ss})}{P_{21}+P_{12}},
\label{eq:}
\end{eqnarray}
where  $\Gamma_{}=P_{21}+P_{12}+\Gamma_{x}'+\Gamma_c$,
and $\Delta_{cx}=\omega_c-\omega_x$. To recover boson
statistics, one simply replaces the $P_{21}+P_{12}$ terms above by $P_{21}-P_{12}$ and sets $\Gamma_x'$  to zero;
 We stress that the
above formulas are  substantially different to previous
models that neglect stimulated emission~\cite{laussyPRL2008,laussyPRB2009,arnePRL2009}; in particular,
we have no unphysical behavior as $\Gamma_c = P_c$,
%we can explore regimes of $P_c>\Gamma_c$,
and we get qualitatively different  saturation behavior
of the QD exciton.
Similar  incoherent pump models, with pump-induced
 stimulated emission, have also been proposed recently by Ridolfo {\em et al.}~\cite{ridolfo_arxiv},
 though they concentrate exclusively on the model-2 exciton pump and they neglect pure dephasing; thus
 their analytical formula applies only to a  boson system.
% One of the most important features of the analytical formulas above, valid for the weakly nonlinear regime,  is that they do include pure dephasing.
% As we show below, both pure dephasing and

%  [cf.~Figs.~2 and 4].

\begin{figure}[!t]
%\hspace{-0.4cm}
%\centering
\vspace{-0.4cm}
\includegraphics[width=0.45\textwidth]{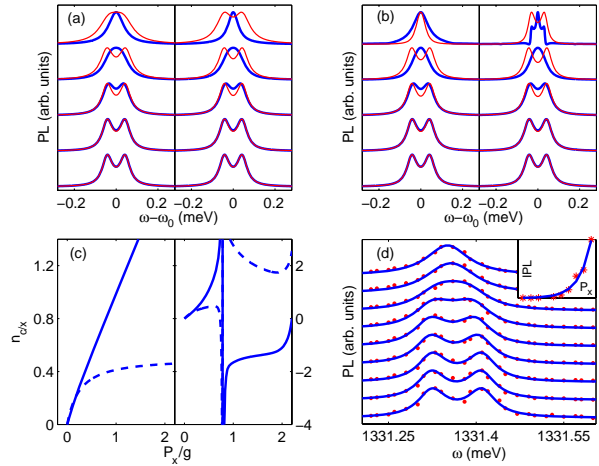}
% Here is how to import EPS art
\caption{(Color online) The on-resonance ($\omega_c \approx \omega_x$)  PL spectra,
%in arbitrary units,
for different excitation powers.
(a) Solution of our  ME with  model 1 (left) and model 2 (right).
The red curve is the one photon spectra and the blue curve is the multi-photon case.
The bottom-to-top panels
%have $P_x = [0.12,4,16,64]\, 1.36g$, and $P_c=1.6P_x$.
have $P_x = [0.12, 0.5, 4, 16, 64]\, 0.02125g\, (0.0003-1.36g)$, and $P_c=1.6P_x$.
(b) ME solution {\em without} stimulated emission.
% with exciton model-1 (left) and model-2 (right).
(c) Mean exciton number (dashed) and photon number (solid),
for our ME with model 1 (left); ME solution {\em without} stimulated emission using
model 1 (right).
%is shown the the right panel.
(d) Experimental data corresponding to
$P_{\rm exp}=[0.12, 0.25, 0.5, 2, 4, 8, 16, 32, 64]\, \mu$W, and
 %our
 model-1 fits (multi-photon and stimulated emission included),
where $P_{x/c}$ proportionally follows the
experimental values; the inset shows the
integrated PL (experiment and theory).}
\label{fig:powerseries}
\end{figure}

{\em The power dependent PL.--} To  highlight the underlying physics of pump-induced PL,
we  proportionally change $P_x$ (and $P_c$) in our model, and keep all other parameters fixed (i.e., $g$, $\Gamma_x$,
 $\Gamma_x'$). The  fixed parameters
 are either known for our experimental system, e.g. $\Gamma_x=0.002$\,meV~\cite{stephanMeasured},
 or are accurately obtained from the
 fitting the experimental data at low powers,
where   $g= 0.045$\,meV and  $\Gamma_c=0.08$\,meV.
We  have also included a dominant pure dephasing exciton decay, $\Gamma_{x}'=0.035$\,meV, caused
 by electron-phonon scattering and spectral diffusion.
The chosen values of $P_x$ range from $0.003-1.36\,g$,
and $P_c=1.6\,P_x$. The justification for allowing $P_c$
to also follow the power of the laser is due to the fact that our micropillar measurements
show a clear linear dependence with power for the cavity mode.
% PL
%cavity mode contribution follows this behavior for all measured powers.
For other QD-cavity systems, such as for a few QDs in a photonic crystal cavity,
$P_c$ may  saturate at much lower powers.
In Fig.~2(a), we  first show the power-dependent spectra
for model-1 (left) and model-2 (right); and  in Fig.~2(b), we compare the trend
expected from a ME model that neglects stimulated
emission processes.
The red
curves show the one photon results and the blue curves show the multi-photon
case.
%\cite{4photons}.
Although all figures show a similar trend of the doublet becoming
a singlet as a function of power, the high power linewidths are substantially
different. In particular, the model with stimulated emission predicts a much
larger pump-induced broadening as a function of power.
In the absence of stimulated emission,
the pump-induced broadening is suppressed,
and the larger pump rates  result in negative
exciton and photon densities.
The mean exciton number (dashed) and photon  number (solid)
 are shown in Fig.~2(c) using the multi-photon model. Here we see the drastic influence
 on the predicted densities if stimulated emission is not included (right), where
 negative photon densities are predicted in addition to regimes
 of $n_x>1$, both of which are obviously unphysical; though we model-1 densities, model-2 gives
 similar unphysical results~\cite{refMun}.
 Of course,
 with stimulated emission neglected in the model, the regime
 of $P_c>\Gamma_c$ is phenomenologically not allowed~\cite{laussyPRB2009}, so the top spectra  in Fig.~2(b) are
 not reliable.
 It is interesting to note
 that, even for pump rates
 as small at $P_x=0.085g$,  multi-photon
 states (c.f.~Jaynes-Cummings model~\cite{carmichael_book})
 are already important contributions to the nonlinear spectra,
 %However, the one-photon spectra can also get a good qualitative agreement with the
 %experim
 and we find that 2-4 photon states
 are enough to get good convergence.
 % in the nonlinear regime shown.
%We do confirm, however, that previous works that
% do not directly include stimulated emission in the ME, namely
% the spectra presented in Ref.~\cite{arnePRL2009}, has indeed net positive densities, for their chosen parameters. But this
% should not distract
% from the main message: high pump powers require stimulated emission  in the model.

 \begin{figure}[t]
%\hspace{-0.4cm}
%\includegraphics[width=0.46\textwidth, height=0.30\textwidth]{fig3.eps}
\includegraphics[width=0.45\textwidth]{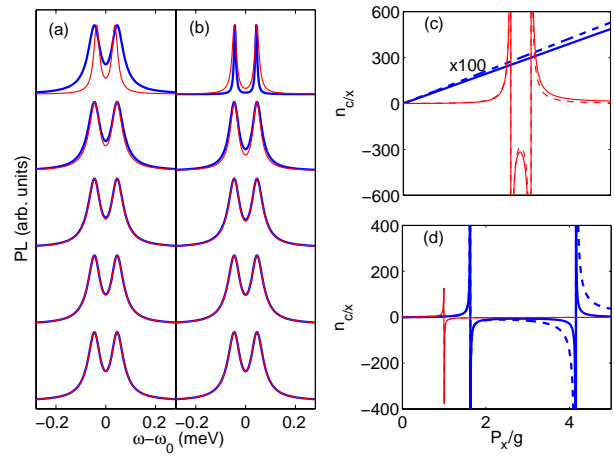}
\vspace{-0.3cm}
% Here is how to import EPS art
\caption{(color online)
The on-resonance ($\omega_c \approx \omega_x$)  PL spectra,
%in arbitrary units,
for different excitation powers,
but for a boson model. (a) ME with stimulated emission, using
exciton pump model-1 (blue) and model-2 (red). (c)
Corresponding mean density plots: exciton number (dashed) and photon numbers (solid); for clarity the model-1 densities are multiplied by 100.
(b,d) As in (a,c), but {\em without} stimulated emission.}
\label{fig:mpn}
\end{figure}

 The experimental data is shown in Fig.~2(d), alongside
 our fermion model-1,
 and there is an excellent correspondence. We stress that the only fitting parameter is  a
 proportionality constant.
%Importantly,  our chosen pump rates
% exactly (proportionally) follow the experimental powers.
Although $\Gamma_x'$ may also be pump-dependent, we find that increasing its value by
 1-2 orders of magnitude has little influence on our high-power PL, as the stimulated--emission-induced
 broadening is by far the dominant source of broadening.
 %Another consequence is that
 %our predicted mean photon numbers are larger, and already
 %in the regime to see exciton lasing, namely $n_c>1$.
 %We stress that
 To have further confidence in the theory,
 %It is
 %very
 it is important that the  models consistently fit the normalized PL,
 on and off-resonance, as well as the integrated PL.
 %In this regard,
 %Indeed,
 We
 %highlight
 %that
 %we
 obtain
 %excellent
 very good
 fits to the spectra
  when the cavity and exciton are off resonance (not shown)
 and the integrated PL [shown as an inset in Fig~2(d)], without changing any
 %fitting
 parameters.

Since our QDs are rather large, e.g., elongated with lengths on the order of 100~nm
and widths of about 30~nm~\cite{reithmaier_nature432}, it is natural to present the nonlinear boson PL
 calculations as well.
In Fig. 3 we display the exact boson PL using exciton-pump models 1 and 2, again with and without
stimulated emission terms. Since pure dephasing cannot be included,
we set $\Gamma_x\rightarrow \Gamma_x+\Gamma_x'$. Clearly, none of the PL follow the trends of the
experiments, and only the thermal bath models
produce net positive densities for all pump rates. Moreover,
even the low PL have different lineshapes due to the important effect of pure
dephasing, which acts to suppress the Rabi oscillations without affecting
the envelope of the population decay.
While it has been discussed before
that the boson model (with model-2)~\cite{munch_oe12821} apparently
fits well to the same data under variation of the
coupling constant g and three other free parameters
($\Gamma_x, P_x, P_c$); we believe that having so many
free parameters can be detrimental to highlighting the underlying
physics.
% (e.g., fermion behaviour as apposed to boson behaviour), and a deliberately constrained model such as ours is
%%clearly
%a much and more satisfying better way to connect to the
%%physics of the
%pump-dependent PL.
We conclude that our nonlinear PL spectra {\em unambiguously} follow
the presented fermion model, and we are thus well into the regime
of anharmonic cavity QED.

%
%%We conclude that our QD-cavity system shows all the signs
%%of anharmonic cavity-QED.
% While it has been discussed before that the boson model (with model-2)~\cite{munch_oe12821}
%  apparently fits well to the same data, we wish to clarify
% here that this was
%due to the fact that  one can usually
%always determine a good fit if $g,\Gamma_x,P_x,P_c$ ({\em four free fitting parameters}) are allowed to
% vary
% to determine a best-fit lineshape; for example, in previously fitting the same
% data~\cite{munch_oe12821}, $g=44-7\,\mu$eV, $P_c=1-8\,\mu$eV,
% $\Gamma_x=37-88\,\mu$eV, and $P_x=1-7\,\mu$eV.
% %Unfortunately,
% We believe that having so many free parameters is detrimental to highlighting the
% correct physics, and a constrained model such as ours is
% %clearly
% a clearer  way to connect of  pump-dependent PL.

\begin{figure}[!t]
%\hspace{-0.4cm}
\includegraphics[width=0.45\textwidth]{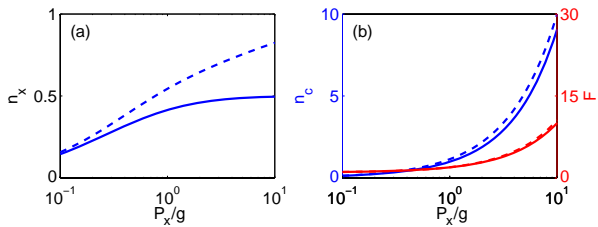}
% Here is how to import EPS art
\vspace{-0.3cm}
\caption{(color online)
(a) Mean exciton number versus
$P_x$ (with $P_c=1.6 P_x$, as before): exciton pump model-1 (solid) and model-2 (dashed).
(b) Corresponding
mean photon number (left axis: blue) and Fano factor $F$ (right axis: red);
%(c) Mean photon number
%with $P_c=0$, using
%$\Gamma_c^{\rm eff}=0.08\,$meV (solid)
%and $\Gamma_c^{\rm eff}=0.01\,$meV (dashed);
%(d) Corresponding mean photon number
%and $g^{(2)}$; the red circles label the laser thresholds
%(see text). All calculations
%use the exact fermion case.
}
\label{fig:detuning}
\end{figure}

{\em High pump-power inversion and lasing--}
Finally, we briefly connect to the prospects for observing
one exciton lasing in such a QD system.
It is well known
in the field of atomic optics, e.g., see
%as discussed by L\"offlet {\em et al}
Ref.~\cite{LofflerPRA1997},
that the spectral properties of pump-dependent PL can be investigated
to explore the regime of single atom lasing.
Characteristic signatures of single state lasing
in atomic physics include spectral narrowing,
inversion,
and a regime of linearly increasing mean photon number as a function
of pump power.
%In fact we have all three of these.
% which is followed by a maximum and a reduction
%on the photon number (panel (b)).
% deduction peak and a maximum, which can be interpreted as a threshold.
On the other hand,  an incoherent pump of thermal photons
will naturally be detrimental to the prospect of achieving
single photon lasing.
%To show this more clearly,
In  Fig.~4(a) we use model-1 (solid) and model-2 (dashed) to investigate
the pump-dependent
 mean exciton number and the mean photon number (panel (b)).
As expected model-1 (thermal bath) does not exhibit any inversion, though model-2
does allow inversion. Both  models allow a mean photon number of greater than 1.
However, the Fano function (photon number variance)~\cite{Fano} shows no evidence of a maximum, and thus there is no lasing threshold in this system.
To achieve single exciton lasing with the present model/system, we have numerically verified that one requires
a much smaller $P_c/P_x$ ratio and a significantly smaller $\Gamma_c$; for example,
$P_c=0$ and $\Gamma_c=0.01\,$meV gives a very clear lasing threshold and order-of-magnitude reductions
in the PL linewidth.  Experimental activity
on single QD lasers has  begun~\cite{QDLasing}, and,
in future work, we will
%future work
%we
explore the
% prospects and
 %experimental
 key signatures
of single exciton lasing using a more detailed multi-level excitation scheme.

{\em Conclusions.--}
A master equation formalism, with incoherent
pumping, pure dephasing, and
a QD fermion model, has been introduced and used to investigate the
 power-dependent PL spectrum of a QD exciton
 under steady-state pumping.
We have shown the importance of self-consistently including
stimulated emission,
% or else negative densities can be predicted.
%
%pointed out the inconsistencies of previous theoretical formalisms
%that neglect pump-induced stimulated emission processes.
%We also have
and validated
our model by directly comparing with recent experimental
data on semiconductor micropillar-cavities.
%, and find that
Using the proposed thermal bath model, an excellent
%%good
fit to the data is obtained  by {\em only} changing the
pump rates in direct correspondence with the experiments, showing that
 {\em we are well into the elusive regime of anharmonic cavity-QED}.
Moreover, we have shown that our excitation models produce positive-definite densities for all pump rates.
%Finally, we have  clarified the drastic differences between
%a fermion and boson model, and briefly connected to the phenomenon
%of single exciton lasing, a regime
%that was possibly approached just recently~\cite{QDLasing}.
%%Future work will explore a more quantitative multi-level excitations scheme.

This work was supported by the National Sciences and Engineering Research Council of Canada,
the Canadian Foundation for Innovation, and the Deutsche Forschungsgemeinschaft via the Research Group {\em Quantum Optics
in Semiconductor Nanostructures} and the State of Bavaria. We thank  H. Carmichael,
J.   Finley, and
A. Laucht  for
 comments.

\end{document}